\documentclass[10pt]{article}
\usepackage{amsmath,amsfonts}
\usepackage{graphicx}
\usepackage{epsfig,epsf}
\usepackage{hyperref}
\usepackage{bm}

\def\bar {\overline}

\def\be {\begin{equation}}
\def\ee {\end{equation}}
\def\beq {\begin{equation}}
\def\eeq {\end{equation}}
\def\bea {\begin{eqnarray}}
\def\eea {\end{eqnarray}}

\def\gh{\Gamma_h}
\def\beq{\begin{equation}}
\def\eeq{\end{equation}}
\def\barr{\begin{array}}
\def\earr{\end{array}}

\begin{document}

\renewcommand*{\thefootnote}{\fnsymbol{footnote}}

\begin{center}
 {\Large\bf{Lepton flavour violating Higgs boson decay $\bm{h\to\mu\tau}$ at the ILC}}

\vspace{5mm}

Indrani Chakraborty \footnote{indrani300888@gmail.com},
Amitava Datta \footnote{adatta\_ju@yahoo.com}, and 
Anirban Kundu \footnote{anirban.kundu.cu@gmail.com}

\vspace{1mm}
{\em{Department of Physics, University of Calcutta, \\
92 Acharya Prafulla Chandra Road, Kolkata 700009, India
}}

\end{center}
\begin{abstract}
 We study the possible reach of the proposed International Linear Collider (ILC) in exploring the 
 lepton flavour violating (LFV) Higgs boson decay $h\to\mu\tau$. Two generic
 types of models are investigated, both involving an extended scalar sector. For the first type, 
 the rest of the scalars are heavy and beyond the reach of ILC but the LFV decay occurs through 
 a tiny admixture of the Standard Model (SM) doublet with the heavy degrees of freedom. In the 
 second class, which is more constrained from the existing data, 
 the SM Higgs boson does not have any LFV decay but there are other scalars degenerate 
 with it that gives rise to the LFV signal. We show that ILC can pin down the branching fraction of $h\to\mu\tau$,
 and hence the effective LFV Yukawa coupling, 
 to a very small value, 
 due to the fact that there are signal channels with unlike flavour 
 leptons but no missing energy. It turns out that the low-energy options of the ILC, namely, 
 $\sqrt{s}=250$ or 500 GeV are better to investigate such channels, and the option of beam 
 polarization helps too. At least an order of magnitude improvement is envisaged over the existing limits, and 
 the effective LFV Yukawa coupling can be probed at the level of $10^{-4}$.

\end{abstract}

\date{\today}




\setcounter{footnote}{0}
\renewcommand*{\thefootnote}{\arabic{footnote}}


\section{Introduction}

The leptonic flavour violating (LFV) decay of the Higgs boson $h \to\mu\tau$, which is a 
smoking gun for physics beyond the Standard Model (SM), has been looked for by 
both CMS and ATLAS collaborations at the Large Hadron Collider (LHC), 
and the branching ratio (BR) is \cite{cms-lfv,atlas-lfv}
\beq
{\rm BR}(h\to\mu\tau) = 0.84^{+0.39}_{-0.37}\%~~{\rm (CMS)}\,,\ \ 
0.53 \pm 0.51\% ~~{\rm (ATLAS)}\,,
\eeq
so that the 95\% CL upper limits on the BR is 1.51\% (CMS) and 1.41\% (ATLAS). 
While there is no official averaging, one might take ${\rm BR}(h\to \mu\tau) < 1.4\%$. 

This has given rise to a lot of model building efforts to explain the signal \cite{theory,thdm3,dipankar,kamenik,sher}. 
As the $\mu$-$\tau$ pair has an invariant mass suggestive of a parent of mass close to that of the Higgs boson, 
there are only two possibilities: (i) By some mechanism (possibly from mixing with other scalars) 
the SM Higgs boson has acquired some small flavour-changing (FC) coupling; (ii) The leptons are 
coming from some other scalar degenerate with the Higgs boson. Both possibilities have been explored 
in the literature \cite{thdm3,dipankar,kamenik,sher}, although the second possibility appears to be 
more constrained from the data. There is also a third possibility of introducing 
the LFV coupling through loop effects without enhancing the SM scalar sector (but other new degrees of 
freedom may be needed) \cite{pilaftsis}; however, 
this is equivalent to the introduction of an effective LFV coupling, as is the case for the first 
possibility. 

Let us assume that the decay occurs through a term $\left[-y_{ij} \bar\ell^i_L \ell^j_R h + {\rm h.c.}
\right]$ in the Lagrangian, where $h$ is the scalar resonance at 125 GeV. Written in full, the relevant term is 
\be
-y_{\mu\tau} \left(\bar\mu_L \tau_R + \bar\tau_R \mu_L\right) h  
-y_{\tau\mu} \left(\bar\tau_L \mu_R + \bar\mu_R \tau_L\right) h\,.
\label{lfv-lag}
\ee
The BR is given by
\be
{\rm BR}(h\to\mu\tau) = \frac{m_h}{8\pi\Gamma_h} \left( |y_{\tau\mu}|^2 + |
y_{\mu\tau}|^2\right) \equiv \frac{m_h}{8\pi\Gamma_h} {\cal F}^2\,.
\ee
If $h\to\mu\tau$ (and other possible new decay channels) have a small BR, 
one can write $\Gamma_h \approx \Gamma_h^{\rm SM} = 4.07$ MeV for $m_h=125$ GeV.

The measurement constrains ${\cal F}=\sqrt{|y_{\mu\tau}|^2 + |y_{\tau\mu}|^2}$. Unless we measure 
the $\mu$ or $\tau$ polarizations, no further information about the individual couplings 
can be drawn. The upper bound from ATLAS \cite{atlas-lfv} suggests 
\be
{\cal F } < 3.4\times 10^{-3}
\ee
at 95\% CL. If there is a hierarchy between the two couplings $y_{\mu\tau}$ and
$y_{\tau\mu}$, this is the bound on the larger coupling. If both are equal, the bound on each of them
is $1/\sqrt{2}$ times that on ${\cal F}$. We will show how far ${\cal F}$ can be probed. Its relationship 
with the Yukawa couplings and mixing angles is model-dependent. 

If the new channels have a significant BR, one should use 
\be
\gh = \frac {\Gamma_h^{\rm SM}}{1-{\rm BR}(h\to {\rm new})}\,.
\ee
Note that such a scenario is quite feasible.
Experimentally, $\Gamma_h$ has a strong lower bound but a very weak upper bound, so 
even larger values of ${\cal F}$ can be allowed in principle. The upper limit on $\Gamma_h$ 
from direct measurement is 1.7 GeV \cite{cms-gamma} and from indirect measurement, 22 MeV 
\cite{gamma-indir}. This makes the $1\sigma$ upper bound on ${\cal F}$ to be 0.063 (direct measurement) 
or 0.007 (indirect measurement) approximately, assuming the only new channel to be $h\to\mu\tau$, for which
one may allow a large branching ratio.

We would like to find out how far the FC coupling(s) can be explored at the International 
Linear Collider (ILC) \cite{ilc}. 
While there are such studies in the literature in the context of supersymmetric LFV 
models \cite{kanemura}, we would like to give more general constraints based on the updated ILC 
parameters. 
Almost all the models that try to explain the FC decay invoke an extended scalar 
sector \footnote{LFV couplings of the Higgs boson must identically vanish if the mass matrix 
is proportional to the Yukawa matrix, as the off-diagonal Yukawa interactions --- even if 
effective --- vanish in the stationary basis. With a single Higgs doublet, one needs to invoke 
higher-dimensional operators for this that can spoil the proportionality between the mass and the 
Yukawa matrices.}.
As we have said before, there are generically two classes of models. The first class, 
which we may call the Lonely Higgs (LH) models, has an extended scalar sector, but all the scalars 
except the Higgs boson $h$ are heavy. One or more of these extra degrees of freedom may mix with the SM 
doublet, and that gives rise to a small FC coupling of $h$. If the FC Yukawa coupling of the heavy scalars be $y_H$ 
and the mixing angle with the SM doublet be $\theta$, we expect ${\cal F} \sim y_H\sin\theta$. the precise relation 
between ${\cal F}$ and $\{y_H,\theta\}$ depends on the model chosen and its parameters. 
The LH models have been widely investigated in the literature and include Type-III 2HDM \cite{thdm3}, 
or scalar sectors with discrete symmetries, 
or even composite Higgs models. 
For example, in Ref.\ \cite{thdm3}, it has been shown that in a flavour-changing 2HDM, the effective 
Yukawa coupling of $\bar\tau\mu h$ is a function of the angles $\alpha$ (the mixing angle of the 
two CP-even neutral states), $\beta$ (the mixing angle of the charged scalar states, and the ratio of 
the two vacuum expectation values $v_2/v_1$ is given by $\tan\beta$), and $\theta_R$ (the mixing 
angle of the charged leptons $\mu_R$ and $\tau_R$), apart from the FC Yukawa coupling of the 
second doublet. Since the functional form depends on the model chosen, we will not go into any 
particular model and express all our results in terms of ${\cal F}$ as defined before. 
We will also assume all other scalars of such LH models to be beyond the reach of ILC. 

The second class, which we may call the Degenerate Higgs (DH) models, involves, apart from $h$, 
more scalars at 125 GeV, some (or all) of them having nonzero FC couplings and giving rise to the 
$\mu$-$\tau$ signal. Examples of DH models are in Refs.\ \cite{dipankar,BGLLFV}. As these models 
are not very well investigated but may be interesting from an experimental point of view regarding 
clear signals or possible falsifiability, let us give a very brief outline of one of the DH models 
discussed in \cite{dipankar}. 

In this model, all the neutral scalars are almost degenerate. The SM 
Higgs boson $h$ has only flavour-conserving couplings; the other CP-even neutral scalar $H$ and the CP-odd 
pseudoscalar $A$ can decay to the flavour-violating channels. The approximate degeneracy is forced by the observed signal,
where the $\mu$-$\tau$ invariant mass peaks at $m_h$. 

In the exact alignment limit, with $\alpha$ and $\beta$ defined before,
\be
\vert \alpha-\beta\vert = (2n+1) \frac{\pi}{2}\,,
\ee
which makes the lighter CP-even eigenstate identical in properties with the SM Higgs boson \footnote{ 
In the other alignment limit $|\alpha-\beta| = n\pi$, the heavy CP-even eigenstate is identical to the 
SM Higgs. This possibility is also not ruled out from the LHC data.}. This is the limit that is favored 
by the LHC data, but small deviations can still be entertained. The amount of the allowed deviation depends 
on the type of 2HDM \cite{alignment}. 

The pseudoscalar $A$ does not have any $WWA$ or $ZZA$ coupling, so it can never be produced by the Bjorken 
process or gauge boson fusion \footnote{At an integrated luminosity of 500 fb$^{-1}$, double scalar productions 
from the two gauge-two scalar quartic interactions are anyway negligibly small.}.  If one is away from the 
alignment limit, the $h$ production rate is $\sigma(h_{SM})\times \sin^2(\beta-\alpha)$, and the $H$ 
production rate is $\sigma(h_{SM})\times \cos^2(\beta-\alpha)$, where $\sigma(h_{SM})$ is the 
SM Higgs boson production cross-section. On the other hand, the FC coupling of $h$ ($H$) is proportional 
to $\cos(\sin)(\beta-\alpha)$. Thus, ${\cal F} \sim y\sin 2(\beta-\alpha)$ where $y$ is the (effective) FC Yukawa coupling
of $H$, like $y_{\mu\tau}$ or $y_{\tau\mu}$ as shown in Eq.\ (\ref{lfv-lag}). 
One must mention that the simplest version of this model as in Ref.\ \cite{dipankar} does not work 
because the model predicts a large branching ratio for $h\to e\tau$ which has not been observed by the 
CMS collaboration \cite{cms-etau}. Still, one may tweak the model and have a deviation from the strict 
alignment limit to satisfy all the experimental constraints \cite{sher}.

Thus, what one can bound from the data is either (i) the Yukawa coupling of the $\mu$-$\tau$ pair with the heavy 
scalar, multiplied by $\sin\theta$, where $\theta$ is the mixing 
angle between the SM Higgs and the heavy neutral one, for the LH models, or
(ii) the Yukawa coupling of the other neutral scalar $H$ to 
$\mu\tau$ multiplied by a measure of the deviation from the 
alignment limit, for the DH models. Any further information would require a knowledge 
of the model, as investigated in Ref.\ \cite{kamenik}.  
Various LFV models were also studied in the literature before this signal came into prominence; for example, we refer 
the reader to Ref.\ \cite{lfv-other}.

\section{The ILC parameters}

The designed CM energies for the ILC are 250 and 500 GeV. While it can be upgraded to 1 TeV in future, we will 
work with the two designed energies only. The reason is that the sensitivity to LFV couplings are best at the low energy 
options. 
ILC is planned to run with four polarization options: P1 $(-0.8,0.3)$, P2 $(0.8,-0.3)$, P3 $(-0.8,-0.3)$ and 
P4 $(0.8,0.3)$ respectively \cite{kawada,ilc-tdr07,1006.3396}. The numbers give the degree of polarization of the beams; for 
example, the first option (P1) means that the $e^-$ beam is 80\% left polarized and the $e^+$ beam is 
30\% right polarized. The effective polarization is given by 
\be
P_{\rm eff} = \frac{P(e^-) + P(e^+)}{1+P(e^-) P(e^+)}\,,
\ee
where $P(e^-)$ is positive for left polarized electron beam and $P(e^+)$ is positive for right polarized 
positron beam. The projected integrated luminosities for the four polarization options at $\sqrt{s} = 250$ GeV, are (in fb$^{-1}$): 
\be
{\rm P1}:  337.5\,,\ \ {\rm P2}: 112.5\,,\ \ {\rm P3}:25\,,\ \ {\rm P4}: 25\,,
\ee
which will finally be upgraded to $1350,450,100,100$ fb$^{-1}$ respectively. At $\sqrt{s} = 500$ GeV, initial (upgraded) integrated luminosities are (in fb$^{-1}$):
\be
{\rm P1}:  200 (1600)\,,\ \ {\rm P2}: 200 (1600)\,,\ \ {\rm P3}:50 (400)\,,\ \ {\rm P4}: 50 (400)\,.
\ee

%
%

\section{Signal and background}

The dominant Higgs production processes at the ILC are the Bjorken process $e^+e^-\to Zh$ and the vector boson 
fusion (VBF) $e^+e^- \to h \nu_e \bar\nu_e (h e^+e^-)$ through $WW (ZZ)$ fusion. 
The cross-section for the former falls with $\sqrt{s}$ and that of the latter rises. At $\sqrt{s}=250$ GeV, 
the Bjorken process dominates, while at $\sqrt{s}=500$ GeV, both the processes become compatible. 

The final state that we look for is $q\bar{q}\mu\tau$ or $\ell^+\ell^-\mu\tau$, with $\ell=e,\mu$. 
To reliably reconstruct the 
$\tau$ from its one-prong or three-prong decays, one should use the collinear approximation, {\em i.e.}, 
the $\tau$ is so boosted that the neutrino goes essentially in the same direction as its parent. If there 
are no other sources of missing energy, this reconstruction is quite reliable \cite{kawada}. Thus, we 
will not consider the $WW$ fusion further, first because the contributions are less than those coming 
from the Bjorken process, and second, because the extra neutrinos make the $\tau$ reconstruction more
uncertain. It is unlikely that to get the LFV signal from a 125 GeV resonance, one has to wait till the 
1 TeV upgrade, when the initial options, through the Bjorken process, provide a better sensitivity 
\footnote{There might be heavier scalars with LFV couplings too, and for them the high-$\sqrt{s}$ option 
is a must, either at the ILC or the $e^+e^-$ mode of the Future Circular Collider (FCC-ee).}. 
If we focus only on the Bjorken process, there are no further sources of missing energy 
and the $\tau$ reconstruction efficiency $\epsilon_\tau$ is about 70\% \cite{kawada}. In the clean environment of the 
$e^+e^-$ collider, the two jets can be detected with almost 100\% efficiency (we do not need to tag the flavour), 
and so can be the muon. The electron detection efficiency is about $89\%$, while that of the muon is 
almost 100\%.
Note that the $\mu\mu\mu\tau$ channel involves an extra combinatoric, {\em i.e.} one needs to check 
which muon pair came from the $Z$ and which one came from the Higgs boson. 
This reduces the efficiency of the three-muon channel by about 20\% compared to the $ee\mu\tau$ channel. which
approximately offsets the gain coming from the better detection efficiency. A detailed discussion will follow. 

There is, of course, one possible source of background as discussed in Ref.\ \cite{kanemura}. This comes from 
the decay $h\to\tau^+\tau^-$, with one of the tau leptons decaying into a muon. If the missing energy is small, 
this can fake the signal event. We have calculated the number of such fake events, $N_{\rm fake}$, for different 
ILC configurations and final states. 
An explicit evaluation (at parton level) was performed by CalcHEP, generating the $q\bar{q} (e^+e^-) 
\mu\tau\nu_\mu\nu_\tau$ final state at the ILC, including all possible amplitudes. We then applied the cuts as mentioned later:
the $q\bar{q} (e^+e^-)$ invariant mass at an interval of $m_Z\pm 10$ GeV, the $\mu\tau$ invariant mass within 
$m_h\pm 10$ GeV, and missing energy less than 20 GeV. The cuts reduce the uncut cross-section by at least 3 orders 
of magnitude. This is expected as the $\mu\tau$ pair can come from many sources, not necessarily from $h\to\tau\tau$. 
After multiplication by the $\tau$ detection efficiency $\epsilon_\tau = 0.7$ and the projected luminosity, the number of 
fake events becomes completely negligible. Slightly conservative estimate is given in Ref.\ \cite{kanemura} which we show in the 
tables.     
While the number of such fake events $N_{\rm fake}$ is completely negligible for $e^+e^-\mu\tau$ final states, 
it is quite small (and even zero for some configurations) for $q\bar{q}\mu\tau$ final states too. 
One of the reasons 
for this is that the number of signal as well as parent background (with $h\to\tau^+\tau^-$) events are much smaller 
for the former case compared to the latter, coming from ${\rm BR}(Z\to e^+e^-)/{\rm BR}(Z\to q\bar q)$. The second 
reason is that the tau lepton can be more easily reconstructed in the former case than the latter.

The cross-section for the Bjorken process and the subdominant $ZZ$ fusion 
are shown in Table \ref{tab:hprod}.

\begin{table}[htbp]
\begin{center}
\begin{tabular}{||c|c||c|c||}
\hline
 $\sqrt{s}$ & Polari- & \multicolumn{2}{c|} {Cross-section (fb)} \\
 (GeV)      & zation & Bjorken  & $ZZ$ fusion \\
 \hline
 &&& \\
 250 & P1 & 333 & 0.81 \\
     & P2 & 227  & 0.52 \\
     & P3 & 196 & 0.75 \\
     & P4 & 148 & 0.62 \\
     &&&\\
 500 & P1 & 105 & 8.62 \\
     & P2 & 71.4 & 5.48 \\
     & P3 & 61.6 & 7.66 \\
     & P4 & 46.4 & 6.25 \\
     &&&\\
\hline
\end{tabular}
\end{center}
\caption{Higgs production cross-section for different $\sqrt{s}$ and polarization options 
at the ILC. Computed with CalcHEP v3.6.23 \cite{calchep}
with ISR and beamstrahlung options switched on. 
}
\label{tab:hprod}
\end{table}

We will consider the following signal processes.

\begin{enumerate}
 \item $e^+e^- \to Zh$, $Z \to q\bar q$, $h \to \mu\tau$: This is the most efficient channel to 
 probe the FC coupling(s) of the SM Higgs boson in the early (low-$\sqrt{s}$) runs of the ILC. 
 The channel has almost zero SM background if we use the cuts mentioned above ---
 we take the invariant mass of the two jets in the interval $m_Z\pm 10$ GeV, the invariant mass of the 
 $\mu$-$\tau$ pair in the interval $m_h \pm 10$ GeV, and missing energy below 20 GeV.
 With these cuts, 
 one retains 90\% or more of the signal events while the backgrounds 
 coming from $t\bar{t}$ or $W^+W^-Z/\gamma$ final states can be successfully eliminated
 \footnote{$t\bar{t}$ final states can give rise to two $b$-jets, $\mu^\pm\tau^\mp$, and missing energy.
 For $W^+W^-Z/\gamma$, one gets the same background (with missing energy) when the two $W$'s decay 
 leptonically, and the $Z/\gamma$ goes to $q\bar{q}$.}. 
 Note that any SM process with two different flavoured leptons must involve at least two neutrinos 
 and therefore a significant missing energy, unless the neutrinos conspire otherwise.

 \item $e^+e^- \to Zh$, $Z\to \ell^+\ell^-$, $h \to \mu\tau$, with $\ell=e,\mu$: The reach is 
 significantly lower than $Z\to q\bar{q}$ because of the BR suppression for $Z\to \ell^+\ell^-$. 
 For $Z\to\mu^+\mu^-$, one has to apply combinatorics for muons in the final state to determine which 
 of the two like-sign muons came from $h$ and which from $Z$. 
 Consider, for example, the final state $\mu^+\mu^-\mu^+\tau^-$. 
 For the event to be acceptable, we have to know for sure which $\mu^+$ originated from $Z$ and 
 this is not possible if 
 the invariant masses of the $\mu^-$ with both $\mu^+$ fall close to the $Z$-peak, say 
 in the $m_Z\pm 5$ GeV range. We have checked that 
 this amounts to a reduction of 20\% at the most. At the same time, the detection efficiency of $\mu$ is 
 almost 100\% compared to 89\% for the electron; which means that one is likely to detect equal number of 
 $ee\mu\tau$ and $\mu\mu\mu\tau$ events. 
 There is a small contribution to $ee\mu\tau$ final state from $ZZ$ fusion which is absent for the 
 three-muon final state. 
 Again, the cuts remove all the backgrounds but never more than 10\% of the signal
 events. We found $N_{\rm fake}$ to be always zero for this case.

\end{enumerate}

Our results are shown in the following tables. We take the integrated luminosity as envisaged by the ILC, and the 
detection efficiencies as mentioned before:
\be 
\epsilon_\tau = 0.7\,,\ \ 
\epsilon_\mu = \epsilon_{\rm jet} = 1.0\,,\ \ 
\epsilon_e = 0.89\,,
\label{effs}
\ee
and the branching ratios
\be
{\rm BR}(Z\to q\bar{q}) =0.7\,,\ \ 
{\rm BR}(Z\to \nu\bar\nu) = 0.2\,, \ \ 
{\rm BR}(Z\to \ell^+ \ell^-)= 0.03363\,.
\label{brs}
\ee
Table \ref{tab:sigevent} shows 
the number of events for various $\sqrt{s}$ and polarization options if ${\rm BR}(h\to\mu\tau)=0.01$, 
where we have taken into account the charge-conjugated final state too. The cuts as mentioned above 
have been applied. From the large number of events 
in channels with zero or small background, it is clear that ILC can probe to a far smaller BR. 

\begin{table}[htbp]
\begin{center}
  \begin{tabular}{ || c | c | c | c | c | c ||}
    \hline
   & $\sqrt{s}$ (GeV) & Polarization &  
 \vtop{\hbox{\strut $Zh$\,,}\hbox{\strut $Z\rightarrow q \bar{q}$}} & 
 \vtop{\hbox{\strut $Zh$\,,}\hbox{\strut $Z\rightarrow e^+ e^-$}} &  
 \vtop{\hbox{\strut $N_{\rm fake}$}} \\ \hline
 Initial & 250  & \vtop{\hbox{\strut P1}\hbox{\strut P2}\hbox{\strut P3}\hbox{\strut P4}}& 
 \vtop{\hbox{\strut 496}\hbox{\strut 113}\hbox{\strut 22}\hbox{\strut 16}} & 
 \vtop{\hbox{\strut 23}\hbox{\strut 5}\hbox{\strut 1}\hbox{\strut 1}} &
  \vtop{\hbox{\strut 2}\hbox{\strut 0}\hbox{\strut 0}\hbox{\strut 0}} \\ \hline
 Final & 250  & \vtop{\hbox{\strut P1}\hbox{\strut P2}\hbox{\strut P3}\hbox{\strut P4}}& 
 \vtop{\hbox{\strut 1983}\hbox{\strut 450}\hbox{\strut 86}\hbox{\strut 65}} &
 \vtop{\hbox{\strut 92}\hbox{\strut 21}\hbox{\strut 5}\hbox{\strut 3}} & 
  \vtop{\hbox{\strut 8}\hbox{\strut 2}\hbox{\strut 0}\hbox{\strut 0}} \\ \hline
Initial & 500  & \vtop{\hbox{\strut P1}\hbox{\strut P2}\hbox{\strut P3}\hbox{\strut P4}}& 
\vtop{\hbox{\strut 93}\hbox{\strut 63}\hbox{\strut 13}\hbox{\strut 10}} & 
\vtop{\hbox{\strut 14}\hbox{\strut 10}\hbox{\strut 3}\hbox{\strut 3}} &
 \vtop{\hbox{\strut 0}\hbox{\strut 0}\hbox{\strut 0}\hbox{\strut 0}} \\ \hline
 Final & 500  & \vtop{\hbox{\strut P1}\hbox{\strut P2}\hbox{\strut P3}\hbox{\strut P4}}& 
 \vtop{\hbox{\strut 741}\hbox{\strut 504}\hbox{\strut 109}\hbox{\strut 82}} & 
 \vtop{\hbox{\strut 119}\hbox{\strut 77}\hbox{\strut 24}\hbox{\strut 19}} & 
  \vtop{\hbox{\strut 3}\hbox{\strut 2}\hbox{\strut 0}\hbox{\strut 0}} \\ \hline
  \end{tabular}
\end{center}
\caption{Number of events, after a reduction by 10\% for the cuts applied, 
assuming ${\rm BR}(h\to\mu\tau)=0.01$ for different ${\cal L}$ and polarization 
options as mentioned in the text. $N_{\rm fake}$ denotes the number of fake events 
for $q\bar{q}\mu\tau$ final states. There are no fake events at this luminosity in the 
$ee(\mu\mu)\mu\tau$ channels.
The number of $\mu\mu\mu\tau$ events is approximately the same as the number of $ee\mu\tau$ events 
because of a cancellation between better detection efficiency and combinatoric loss; see text 
for details.} 
  \label{tab:sigevent}
  \end{table}

\begin{table}[htbp]
 \begin{center}
  \begin{tabular}{ || c | c | c | c | c |c | c | c ||}
    \hline
 & $\sqrt{s}$ (GeV) & Polarization & $\mathcal{L}$ (fb$^{-1}$) & ${\rm BR} (5,{\cal L})$  & ${\cal F} (5)$
 & ${\rm BR} (3,{\cal L})$ &  ${\cal F} (3)$ \\ \hline
 Initial & 250  & \vtop{\hbox{\strut P1}\hbox{\strut P2}\hbox{\strut P3}\hbox{\strut P4}}& 
 \vtop{\hbox{\strut 337.5}\hbox{\strut 112.5}\hbox{\strut 25}\hbox{\strut 25}} & 
 \vtop{\hbox{\strut 1.1$\times$10$^{-4}$}\hbox{\strut 3.9$\times$10$^{-4}$}
 \hbox{\strut 2.1$\times$10$^{-3}$}\hbox{\strut 2.7$\times$10$^{-3}$}} 
 & \vtop{\hbox{\strut 2.9$\times$10$^{-4}$}\hbox{\strut 5.7$\times$10$^{-4}$}
 \hbox{\strut 1.3$\times$10$^{-3}$}\hbox{\strut 1.5$\times$10$^{-3}$}} &
 \vtop{\hbox{\strut 7.3$\times$10$^{-5}$}\hbox{\strut 2.4$\times$10$^{-4}$}
 \hbox{\strut 1.2$\times$10$^{-3}$}\hbox{\strut 1.6$\times$10$^{-3}$}} 
 & \vtop{\hbox{\strut 2.4$\times$10$^{-4}$}\hbox{\strut 4.4$\times$10$^{-4}$}
 \hbox{\strut 1.0$\times$10$^{-3}$}\hbox{\strut 1.2$\times$10$^{-3}$}} \\ \hline
 Final & 250  & \vtop{\hbox{\strut P1}\hbox{\strut P2}\hbox{\strut P3}\hbox{\strut P4}}& 
 \vtop{\hbox{\strut 1350}\hbox{\strut 450}\hbox{\strut 100}\hbox{\strut 100}} &
 \vtop{\hbox{\strut 4.1$\times$10$^{-5}$}\hbox{\strut 1.2$\times$10$^{-4}$}
 \hbox{\strut 5.2$\times$10$^{-4}$}\hbox{\strut 6.9$\times$10$^{-4}$}} & 
 \vtop{\hbox{\strut 1.8$\times$10$^{-4}$}\hbox{\strut 3.1$\times$10$^{-4}$}
 \hbox{\strut 6.5$\times$10$^{-4}$}\hbox{\strut 7.5$\times$10$^{-4}$}} &
 \vtop{\hbox{\strut 3.2$\times$10$^{-5}$}\hbox{\strut 7.9$\times$10$^{-5}$}
 \hbox{\strut 3.1$\times$10$^{-4}$}\hbox{\strut 4.1$\times$10$^{-4}$}} 
 & \vtop{\hbox{\strut 1.6$\times$10$^{-4}$}\hbox{\strut 2.5$\times$10$^{-4}$}
 \hbox{\strut 5.0$\times$10$^{-4}$}\hbox{\strut 5.8$\times$10$^{-4}$}} \\ \hline
Initial & 500  & \vtop{\hbox{\strut P1}\hbox{\strut P2}\hbox{\strut P3}\hbox{\strut P4}}& 
\vtop{\hbox{\strut 200}\hbox{\strut 200}\hbox{\strut 50}\hbox{\strut 50}} & 
\vtop{\hbox{\strut 4.8$\times$10$^{-4}$}\hbox{\strut 7.1$\times$10$^{-4}$}
\hbox{\strut 3.3$\times$10$^{-3}$}\hbox{\strut 4.4$\times$10$^{-3}$}} & 
\vtop{\hbox{\strut 6.3$\times$10$^{-4}$}\hbox{\strut 7.6$\times$10$^{-4}$}
\hbox{\strut 1.6$\times$10$^{-3}$}\hbox{\strut 1.9$\times$10$^{-3}$}} &
\vtop{\hbox{\strut 2.9$\times$10$^{-4}$}\hbox{\strut 4.3$\times$10$^{-4}$}
 \hbox{\strut 1.9$\times$10$^{-3}$}\hbox{\strut 2.6$\times$10$^{-3}$}} 
 & \vtop{\hbox{\strut 4.9$\times$10$^{-4}$}\hbox{\strut 5.9$\times$10$^{-4}$}
 \hbox{\strut 1.3$\times$10$^{-3}$}\hbox{\strut 1.5$\times$10$^{-3}$}}   \\ \hline
 Final & 500  & \vtop{\hbox{\strut P1}\hbox{\strut P2}\hbox{\strut P3}\hbox{\strut P4}}& 
 \vtop{\hbox{\strut 1600}\hbox{\strut 1600}\hbox{\strut 400}\hbox{\strut 400}} & 
 \vtop{\hbox{\strut 8.5$\times$10$^{-5}$}\hbox{\strut 1.1$\times$10$^{-4}$}
 \hbox{\strut 4.1$\times$10$^{-4}$}\hbox{\strut 5.5$\times$10$^{-4}$}} & 
 \vtop{\hbox{\strut 2.6$\times$10$^{-4}$}\hbox{\strut 2.9$\times$10$^{-4}$}
 \hbox{\strut 5.8$\times$10$^{-4}$}\hbox{\strut 6.7$\times$10$^{-4}$}} & 
 \vtop{\hbox{\strut 6.1$\times$10$^{-5}$}\hbox{\strut 7.1$\times$10$^{-5}$}
 \hbox{\strut 2.5$\times$10$^{-4}$}\hbox{\strut 3.3$\times$10$^{-4}$}} 
 & \vtop{\hbox{\strut 2.2$\times$10$^{-4}$}\hbox{\strut 2.4$\times$10$^{-4}$}
 \hbox{\strut 4.5$\times$10$^{-4}$}\hbox{\strut 5.2$\times$10$^{-4}$}}\\ \hline     
   
  \end{tabular}
\end{center}
\caption{Reach of ${\cal F}$ in the $\mu\tau q\bar{q}$ channel, for observing 5 events (columns 5 and 6)
or 3 events (columns 7 and 8) at the designed ${\cal L}$ and 
polarization options. This is based on either 5(3) events where the background is zero, or a significance 
of $5\sigma(3\sigma)$ where the background is nonzero.}
\label{tab:qqbar}
\end{table}

For brevity, let us denote the cross-sections for the Bjorken process and $ZZ$ fusion by
$\sigma_b$ and $\sigma_{ZZ}$ respectively. If $N$ events are necessary to claim a 
discovery where the background is negligible, the BR can be probed up to
\be
{\rm BR}(N\,,{\cal L}) = \frac{N}{\sigma_b Z_f \epsilon_f \epsilon_\tau \epsilon_\mu {\cal L}}\,,
\ee
where $Z_f$ is BR of $Z$ in the corresponding fermion channel, as shown in Eq.\ (\ref{brs}), and $\epsilon_f = 
\epsilon_{\rm jet}, \epsilon_e$ and $\epsilon_\mu$ for the $q\bar q$, $e^+ e^-$, and $\mu^+\mu^-$  channels
respectively. ${\cal L}$ is the 
integrated luminosity in fb$^{-1}$. We quote our numbers for $N=5$ and $N=3$ in Table \ref{tab:qqbar}.  
Also shown are the $5\sigma$ and $3\sigma$ ranges where the background coming from fake events are nonzero, 
using the standard $S/\sqrt{S+B}$ measure for the significance. Note that only for the 
$e^+e^-\mu\tau$ final states, $\sigma_b Z_f \epsilon_f$ is to be replaced by 
$\sigma_b Z_f \epsilon_f + \sigma_{ZZ}$. For leptonic final states, we add both $ee\mu\tau$ and $\mu\mu\mu\tau$ 
events to increase the statistics. 

The range of ${\cal F}$ that can be probed is thus given by
\be
{\cal F} = \sqrt{ \frac{8\pi\Gamma_h \, {\rm BR}(N,{\cal L})}{m_h}  }\,.
\ee

The best results are obtained for $\mu\tau q\bar{q}$ final states and for low-$\sqrt{s}$ 
options of ILC. As we see from Table \ref{tab:qqbar}, the 
maximum reach can be obtained for $\sqrt{s}=250$ GeV, and the polarization options also help. 
 With zero background, even 5 signal events mean a discovery, or a $5\sigma$ significance with nonzero 
 background. 
 With the cuts and detection efficiencies as mentioned, a $5\sigma$ significance with 
 the P1 option at $\sqrt{s}=250$ GeV and integrated luminosity 
 ${\cal L}=1350$ fb$^{-1}$ gives ${\rm BR}(h\to\mu\tau) \approx 4.1\times 10^{-5}$, or 
 ${\cal F} \approx 1.8\times 10^{-4}$. This is the best reach of the ILC. Obviously, the reach will 
 be somewhat better with lesser significance.

\begin{figure}[htbp]
\hspace*{-5mm}
\includegraphics[height=6cm,width=8cm]{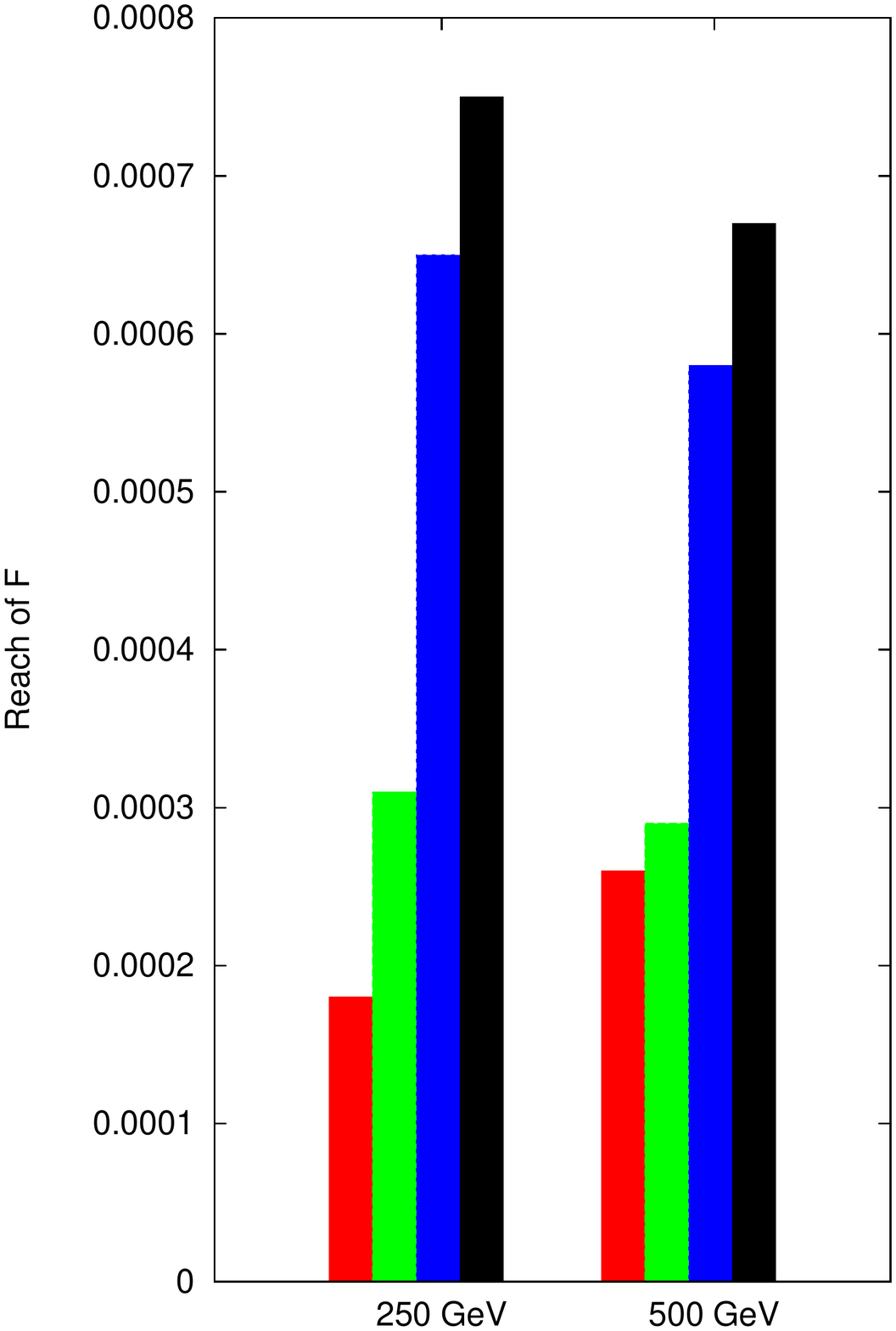} \hspace*{1cm}
\includegraphics[height=6cm,width=7cm]{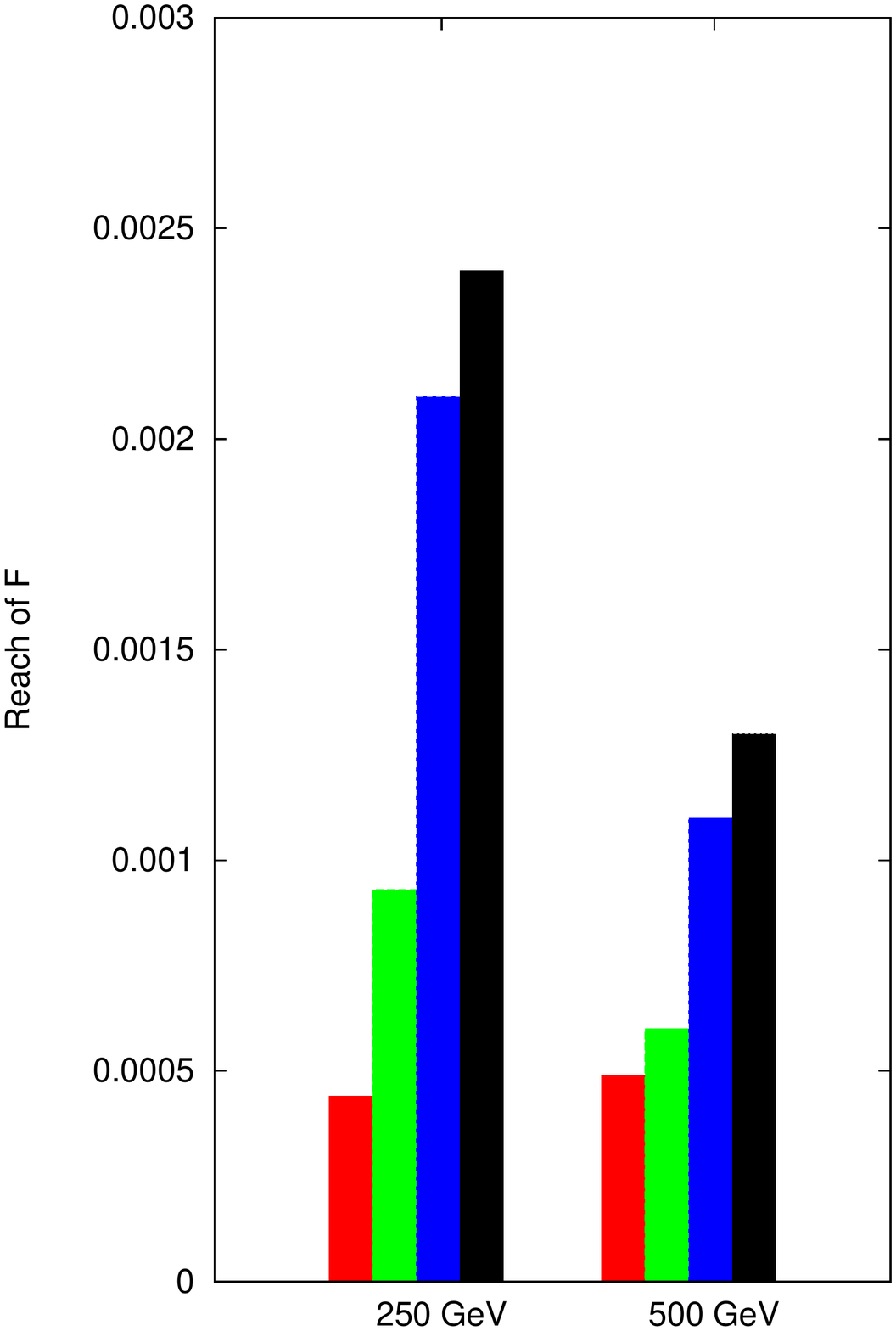}
\caption{The reach for ${\cal F}$ (see text for the definition) for different choices of $\sqrt{s}$,
and with the larger ({\em i.e.}, final) luminosity option. 
The red, green, blue, and black bars are respectively for the polarization options P1, P2, P3 and 
P4 with four different integrated luminosities mentioned earlier. 
The left (right) plot shows the reach for the $\mu\tau q\bar{q}$  
($\mu\tau \ell^+\ell^-$) channel. With nonzero backgrounds, the bars show the $5\sigma$ 
reach; with zero background, the reach with 5 events.}
\label{fig:barchart}
\end{figure}

The reach in the $\ell^+\ell^-\mu\tau$ channel is much worse compared to 
that of $q\bar{q}\mu\tau$, even though 
the former is background-free, because of the lower BR of $Z$ to $\ell^+\ell^-$. 
The results are displayed in 
Table \ref{tab:ee}. A comparative idea of different channels can also be found in Fig.\ \ref{fig:barchart}.

 \begin{table}[htbp]
\begin{center}
  \begin{tabular}{ || c | c | c | c | c |c | c | c ||}
    \hline
& $\sqrt{s}$ (GeV) & Polarization & $\mathcal{L}$ (fb$^{-1}$) & ${\rm BR} (5, \mathcal{L})$& 
${\cal F} (5)$  & ${\rm BR} (3, \mathcal{L})$& ${\cal F} (3)$\\ \hline
 Initial & 250  & \vtop{\hbox{\strut P1}\hbox{\strut P2}\hbox{\strut P3}\hbox{\strut P4}}& 
 \vtop{\hbox{\strut 337.5}\hbox{\strut 112.5}\hbox{\strut 25}\hbox{\strut 25}} & 
 \vtop{\hbox{\strut 9.6$\times$10$^{-4}$}\hbox{\strut 4.2$\times$10$^{-3}$}
 \hbox{\strut 2.2$\times$10$^{-2}$}\hbox{\strut 2.8$\times$10$^{-2}$}} & 
 \vtop{\hbox{\strut 8.9$\times$10$^{-4}$}\hbox{\strut 1.9$\times$10$^{-3}$}
 \hbox{\strut 4.2$\times$10$^{-3}$}\hbox{\strut 4.9$\times$10$^{-3}$}} &
 \vtop{\hbox{\strut 5.8$\times$10$^{-4}$}\hbox{\strut 2.5$\times$10$^{-3}$}
 \hbox{\strut 1.3$\times$10$^{-2}$}\hbox{\strut 1.7$\times$10$^{-2}$}} & 
 \vtop{\hbox{\strut 6.9$\times$10$^{-4}$}\hbox{\strut 1.4$\times$10$^{-3}$}
 \hbox{\strut 3.3$\times$10$^{-3}$}\hbox{\strut 3.8$\times$10$^{-3}$}} \\ \hline
 Final & 250  & \vtop{\hbox{\strut P1}\hbox{\strut P2}\hbox{\strut P3}\hbox{\strut P4}}& 
 \vtop{\hbox{\strut 1350}\hbox{\strut 450}\hbox{\strut 100}\hbox{\strut 100}} &
 \vtop{\hbox{\strut 2.4$\times$10$^{-4}$}\hbox{\strut 1.1$\times$10$^{-3}$}
 \hbox{\strut 5.4$\times$10$^{-3}$}\hbox{\strut 7.1$\times$10$^{-3}$}} & 
 \vtop{\hbox{\strut 4.4$\times$10$^{-4}$}\hbox{\strut 9.3$\times$10$^{-4}$}
 \hbox{\strut 2.1$\times$10$^{-3}$}\hbox{\strut 2.4$\times$10$^{-3}$}} &
 \vtop{\hbox{\strut 1.4$\times$10$^{-4}$}\hbox{\strut 6.4$\times$10$^{-4}$}
 \hbox{\strut 3.2$\times$10$^{-3}$}\hbox{\strut 4.3$\times$10$^{-3}$}} & 
 \vtop{\hbox{\strut 3.4$\times$10$^{-4}$}\hbox{\strut 7.2$\times$10$^{-4}$}
 \hbox{\strut 1.6$\times$10$^{-3}$}\hbox{\strut 1.9$\times$10$^{-3}$}} \\ \hline
Initial & 500  & \vtop{\hbox{\strut P1}\hbox{\strut P2}\hbox{\strut P3}\hbox{\strut P4}}& 
\vtop{\hbox{\strut 200}\hbox{\strut 200}\hbox{\strut 50}\hbox{\strut 50}} & 
\vtop{\hbox{\strut 2.3$\times$10$^{-3}$}\hbox{\strut 3.6$\times$10$^{-3}$}
\hbox{\strut 1.2$\times$10$^{-2}$}\hbox{\strut 1.5$\times$10$^{-2}$}} & 
\vtop{\hbox{\strut 1.4$\times$10$^{-3}$}\hbox{\strut 1.7$\times$10$^{-3}$}
\hbox{\strut 3.2$\times$10$^{-3}$}\hbox{\strut 3.6$\times$10$^{-3}$}} &
\vtop{\hbox{\strut 1.4$\times$10$^{-3}$}\hbox{\strut 2.1$\times$10$^{-3}$}
 \hbox{\strut 7.4$\times$10$^{-3}$}\hbox{\strut 9.3$\times$10$^{-3}$}} & 
 \vtop{\hbox{\strut 1.1$\times$10$^{-3}$}\hbox{\strut 1.3$\times$10$^{-3}$}
 \hbox{\strut 2.5$\times$10$^{-3}$}\hbox{\strut 2.8$\times$10$^{-3}$}} \\ \hline
 Final & 500  & \vtop{\hbox{\strut P1}\hbox{\strut P2}\hbox{\strut P3}\hbox{\strut P4}}& 
 \vtop{\hbox{\strut 1600}\hbox{\strut 1600}\hbox{\strut 400}\hbox{\strut 400}} & 
 \vtop{\hbox{\strut 2.9$\times$10$^{-4}$}\hbox{\strut 4.4$\times$10$^{-4}$}
 \hbox{\strut 1.5$\times$10$^{-3}$}\hbox{\strut 1.9$\times$10$^{-3}$}} & 
 \vtop{\hbox{\strut 4.9$\times$10$^{-4}$}\hbox{\strut 6.0$\times$10$^{-4}$}
 \hbox{\strut 1.1$\times$10$^{-3}$}\hbox{\strut 1.3$\times$10$^{-3}$}} &
 \vtop{\hbox{\strut 1.7$\times$10$^{-4}$}\hbox{\strut 2.7$\times$10$^{-4}$}
 \hbox{\strut 9.2$\times$10$^{-4}$}\hbox{\strut 1.2$\times$10$^{-3}$}} & 
 \vtop{\hbox{\strut 3.8$\times$10$^{-4}$}\hbox{\strut 4.7$\times$10$^{-4}$}
 \hbox{\strut 8.7$\times$10$^{-4}$}\hbox{\strut 9.8$\times$10$^{-4}$}} \\ \hline     
   
  \end{tabular}
\end{center}
\caption{Reach of ${\cal F}$ in the $\mu\tau\ell^+\ell^-$ channel, 
summing over $\ell=e$ and $\ell=\mu$, for observing 5(3) events at 
the designed ${\cal L}$ and polarization options. This channel has negligible $N_{\rm fake}$.
}
\label{tab:ee}
\end{table}


\section{Summary}

In view of the LFV channel $h\to\mu\tau$ with a substantially strong hint at the LHC, 
we investigate the reach of ILC to probe the corresponding LFV Yukawa coupling. 
The advantage of ILC is a relatively clean 
environment and much less QCD background compared to the LHC. An important aspect of the signal is that
any SM background has to contain missing $p_T$ coming from the two 
neutrinos. Thus, if the Higgs boson is produced through the Bjorken process and the $Z$ decays to two jets 
or two leptons, one may have a signal which is background-free. This helps us to achieve at least one 
order of improvement to what we have at the LHC now. The reach for the Yukawa coupling ${\cal F}$ 
also depends on the integrated luminosity ${\cal L}$ and goes as $1/\sqrt{{\cal L}}$.
With an integrated luminosity of ${\cal L}=1350$ fb$^{-1}$, 
${\rm BR}(h\to\mu\tau)$ can be probed up to a level of $4.1\times 10^{-5}$ even at 
$\sqrt{s}=250$ GeV, which in turn means 
 ${\cal F} \approx 1.8\times 10^{-4}$. This is more than an order of magnitude improvement 
 over the LHC. 
The Bjorken process is favoured over the $WW$ fusion because of the absence of missing energy. 
The reach also depends on the various choices of the beam polarization options. 
To compare with the possibility of unpolarized beams, the reach of ${\cal F}$ is $2.1\times 10^{-4}$ 
for identical $\sqrt{s}$ and integrated luminosity, when the polarization option is switched off. 

From the above discussion as well as from the tables, it is obvious that the discovery will most probably 
be made at the low-$\sqrt{s}$ options. This is easy to understand: the Bjorken cross-section falls 
with increasing $\sqrt{s}$, which is nothing but a propagator suppression of the amplitude. 
At $\sqrt{s}=1$ TeV, the 
VBF cross-section becomes important. The 
$WW$ fusion has neutrinos associated with it and therefore one must take care of the background coming from 
$e^+e^-\to W^+W^-$ followed by leptonic decays of the $W$. 

The job of the theoreticians is to translate the model parameters, for both LH and DH types to ${\cal F}$. 
If the new Yukawa couplings are predicted by the model, the experiment will give the mixing angle $\theta$
between the SM doublet and the new scalar. For the DH models, the data will give the shift from the 
alignment limit. 

The ILC will similarly be a good probe for any LFV couplings of the Higgs, like $h\to e\mu$ or $h\to e\tau$. 
The reaches will be quite similar as one looks at the Bjorken process followed by hadronic $Z$ decays. Note that 
the $e\mu$ final state will have a detection efficiency better than final states containg a $\tau$ lepton, 
and so the reach should improve. However, there is no hint for any excess in these channels as yet 
\cite{cms-etau}, and so we do not go into any discussion of them.

{\em Note added:} When the paper was complete, Ref.\ \cite{LFV-Biplob} came to the arXiv which also 
treats the LFV decays at the ILC, also including the possible 1 TeV option.

\centerline{\bf{Acknowledgements}}

I.C.\ acknowledges the Council for Scientific and Industrial Research, Government of India, for a 
research fellowship. A.K.\ acknowledges the Department of Science and Technology, Government of  
India, and the Council for Scientific and Industrial Research, Government of India, for support 
through research grants.

\end{document}